# Fast Sequential Summation Algorithms Using Augmented Data Structures

Vadim Stadnik
vadim.stadnik@gmail.com

**Abstract**

This paper provides an introduction to the design of augmented data structures that offer an efficient representation of a mathematical sequence and fast sequential summation algorithms, which guarantee both logarithmic running time and logarithmic computational cost in terms of the total number of operations. In parallel summation algorithms, logarithmic running time is achieved by high linear computational cost with a linear number of processors. The important practical advantage of the fast sequential summation algorithms is that they do not require supercomputers and can run on the cheapest single processor systems.

**Layout**

1. Motivation
2. Parallel Summation Algorithms
3. Choice of Data Structure
4. Geometric Interpretation
5. Representation of Sequence
6. Fast Sequential Summation Algorithms
7. Mean Value and Standard Deviation
8. Other Data Structures

## 1. Motivation

The original version of the fast sequential summation algorithms described here was implemented in C++ using data structures that support interfaces of STL containers [1]. These interesting and important algorithms can be implemented in many other programming languages. This discussion provides an introduction to the design of data structures, which significantly improve performance of summation algorithms.

The fast summation algorithms have been developed by applying the method of augmenting data structures [2]. This powerful method can be used to solve a wide range of problems. It helps design advanced data structures that support more efficient algorithms than basic data structures. The main disadvantage of this method is its complexity, which makes the use of this method difficult in practice. For this reason, we consider the fast summation algorithms from perspectives that simplify the explanation. The discussion should also help better understand the design principles for quite complex augmented data structures.

Here we focus on the efficient representation of a mathematical sequence and the efficient summation algorithms. For the implementation of update operations on augmented data structures, we refer the reader to the detailed description of the method of augmenting in [2]. To highlight the main ideas, we use simplified variants of the data structures originally implemented in C++.

## 2. Parallel Summation Algorithms

The summation algorithms are fundamental to a wide range of applications from scientific to commercial. The linear running time of the standard sequential summation is acceptable for processing small data sets, but can become a bottleneck for large and huge data sets. Any improvement in the running time of summation algorithms in performance critical applications is of significant importance.

Recent progress in computer technology has enabled the development of effective parallel summation algorithms. Most modern computers have a very small number of processors compared to a typical number of elements processed. The running time of parallel summation on these computers is linear, with optimal improvement over the standard sequential summation proportional to the number of processors. This performance improvement might be significant for some applications, but not acceptable for others. If an algorithm has a performance bottleneck caused by the standard sequential summation, it will not be removed by the parallel summation, which has the same asymptotical running time. The barrier of linear running time can be broken using the ideal parallel machine. The summation algorithm shown in Figure 1 computes the sum in logarithmic time.

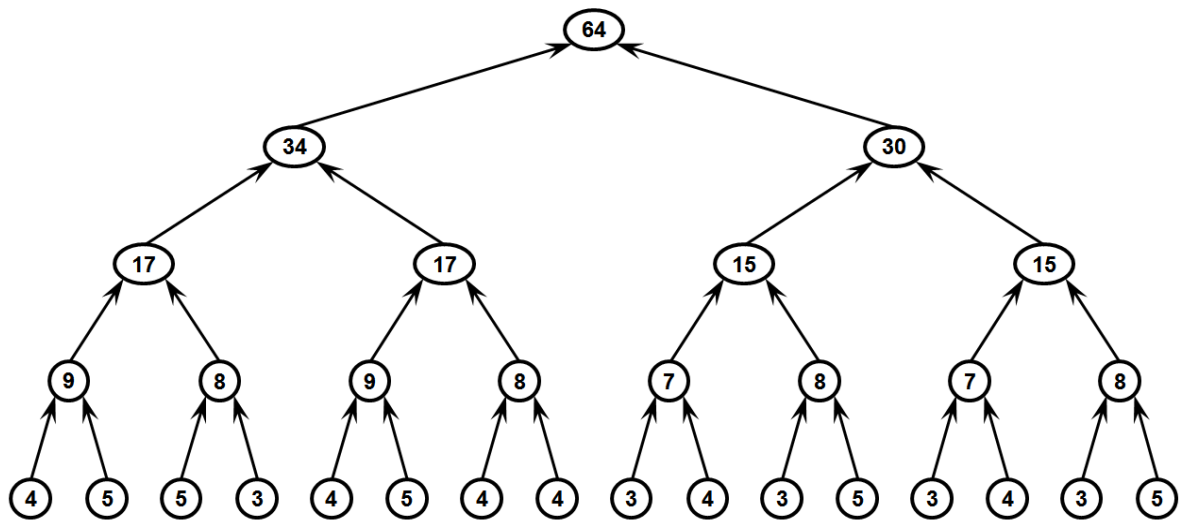

**Figure 1:** A diagram of a parallel summation algorithm for 16 elements. The algorithm takes 4 time steps and performs 15 additions.

A brief discussion of this simple parallel summation algorithm is useful for gaining insight into efficient sequential algorithms.

The ideal parallel summation algorithm achieves considerable performance gain by applying the divide and conquer technique, which avoids sequential processing of a large number of elements that has linear running time. The algorithm uses system processors to divide a given data set into pairs of adjacent elements. The sums of all of the pairs are computed simultaneously and become elements of a new data set to be processed at the next stage of the computations. Each stage of processing decreases the number of elements and involved processors by a factor of 2. The last stage computes the total sum of all of the elements in the given data set. The running time of this algorithm is determined by the number of stages, which is logarithmic in the number of elements.

One of the main features of the parallel summation algorithm is that its logarithmic running time is achieved by high computational cost and, thus, high energy consumption. The total number of operations performed by the parallel algorithm may be even greater than the number of operations performed by the standard sequential algorithm [3]. The drawback of the parallel summation algorithm is that it is designed to compute one value of the total sum of all of the elements in a data set and throws away all of the results computed at intermediate stages. When the processed data set is modified by a small number of operations, such as appending new elements, many intermediate results remain valid and might be reused in subsequent computations. The recomputation of all of the valid data, which is performed by the parallel summation algorithm, incurs additional computational cost and seems to be unnecessary.

This observation leads to the problem of the development of an optimal summation algorithm that performs the minimal number of operations by using precomputed data and avoiding the recomputation of valid data when a data set is modified.

The fast sequential summation algorithms described below provide a solution to this problem. These algorithms guarantee both logarithmic running time and logarithmic computational cost in terms of the total number of operations. They do not require a supercomputer and can run on the cheapest single processor systems, including mobile devices.

## 3. Choice of Data Structure

The implementation of the new summation algorithms requires a data structure that stores and efficiently updates the precomputed data. The choice of the most suitable data structure plays a critical role in the design of such algorithms. In general, the maximum efficiency of an algorithm is limited by the complexity and effectiveness of structural relationships between data items stored in a data structure. For example, both linked lists and balanced trees can store the same data set. A list provides search operations with linear running time only, whereas a tree guarantees significantly better logarithmic running time.

The sequential summation algorithms that operate on linked lists have the same performance characteristics. These basic data structures cannot achieve a running time better than linear. The performance of the sequential summation can be improved using data structures that are more complex than a linked list or an array.

The discussed parallel summation algorithm suggests a data structure that can support the fast sequential summation algorithms. The diagram of the dynamics of parallel computations (see Figure 1) shows the dependencies between the summation operations as a directed acyclic graph. The structural relationships between intermediate results computed by the parallel algorithm at current and next stages are the same as child-parent links in a binary tree. This property of the parallel summation algorithm implies that the implementation of the fast sequential summation algorithms might be based on a binary tree. This is a good option, but not the only one. There are several other data structures suitable for this task.

In the context of the development of the fast sequential summation algorithms, a B+ tree has a number of advantages over other types of trees. One of them is the globally balanced structure. This feature of B+ trees simplifies the design and development of new summation algorithms by using the similarity with the parallel algorithm. A data set stored at one level of a B+ tree can be regarded as a counterpart of a data set computed at one stage of the parallel summation algorithm. The difference between these two data sets is in the method of grouping elements.

A B+ tree that stores precomputed data for summation algorithms can be built in linear time by the following bottom-up algorithm (see Figure 2). The process starts with the construction of a doubly linked list that represents the original numeric sequence. Each next level is built on the top of the previous level. All of the nodes at a new level are linked in a doubly linked list. Each node at a new level is the parent of a group of nodes at the previous level. For illustration purposes, this variant of a B+ tree supports the reduced set of child-parent links: only a parent and its first child are linked. The parent node stores the sum of the values of its child nodes. As the construction proceeds, the number of nodes at a new top level quickly decreases. The process stops when the number of nodes at the top level is less than the maximum size of a group of child nodes. The obtained data structure represents a level-linked B+ tree [4, 5].

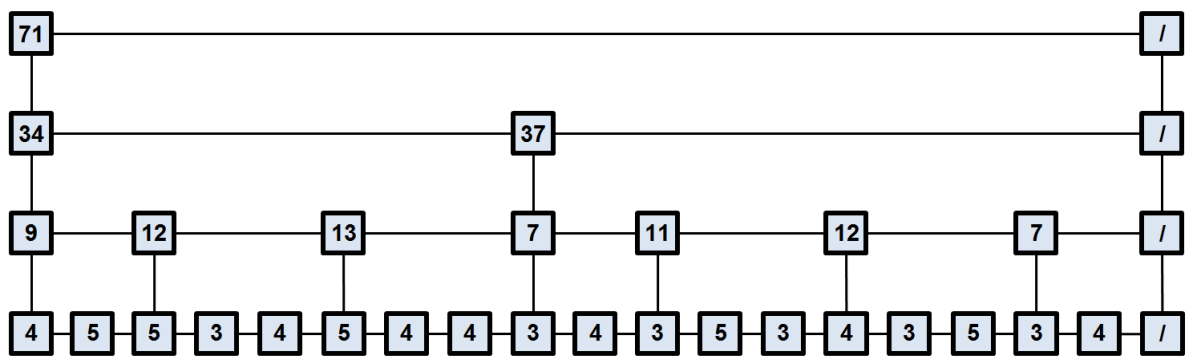

**Figure 2:** A level-linked B+ tree that stores precomputed data for sequential summation algorithms.

## 4. Geometric Interpretation

This variant of a B+ tree offers the advantage of visualizing summation algorithms in one dimensional space (see Figure 3). An element of a sequence of positive numbers can be represented as a line segment with the length equal to the element value. The geometric interpretation combines the representation of a line segment with the representation of the corresponding horizontal link from the current node to the next node. In this data structure, a group of consecutive nodes and a linked list of any level represent a polygonal chain or polyline.

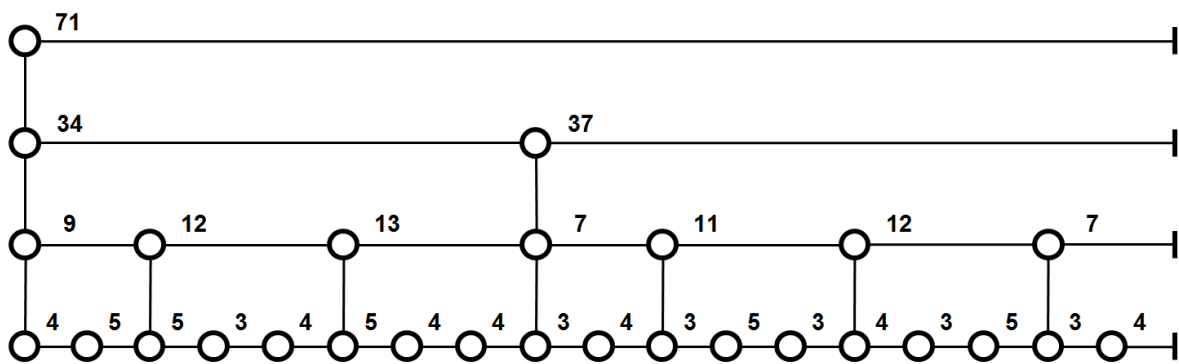

**Figure 3:** A geometric representation of the B+ tree shown in Figure 2. A sum stored in a node in the original tree is mapped to the length of a line segment between the current node and the next node. In a more detailed representation, each horizontal link might be drawn with a length proportional to a corresponding sum. This drawing method is not used here to make the geometric representation more compact.

In contrast to the horizontal dimension, the vertical dimension of the data structure shown does not exist physically. It is only a feature of the schematic representation of this data structure. This dimension is used to separate horizontally linked lists of different levels and to show the child-parent links that connect these levels. The vertical child-parent links are perpendicular to line segments at any level and, thus, their projections on the horizontal dimension do not have lengths.

At any internal level, the start and end points of a parent line segment coincide with the start point of its first child segment and the end point of its last child segment, respectively. The relationship between a parent node and its children can be regarded as a partition of the parent line segment into a polygonal chain of its child line segments.

The deepest level of such a geometric data structure stores line segments that correspond to all of the elements in the original sequence. At all other levels, each line segment has a length equal to the sum of the lengths of its child line segments. When an algorithm moves up, the number of line segments at the current level quickly decreases and the length of each line segment increases. A line segment at an internal level covers a large number of consecutive line segments at the deepest level. The total length of a polyline at each level remains constant, despite the numbers and the lengths of line segments possibly varying quite significantly.

In terms of processing the original sequence, the length of a line segment at an internal level of the geometric data structure represents a sum of a large number of original consecutive elements. The line segment at the top level has a length equal to the total length of all line segments at the deepest level. This result is equivalent to the top node in the original data structure storing the sum of all of the elements.

The geometric data structure of line segments helps explain the following simple rules for the development of new sequential summation algorithms:

1. When an algorithm traverses a horizontal link from the current to the next node (from left to right in Figure 3), it adds the length of the visited line segment. This traversal is equivalent to the addition operation.
2. When an algorithm traverses a horizontal link from the current to the previous node (from right to left in Figure 3), it subtracts the length of the visited line segment. This traversal is equivalent to the subtraction operation.
3. A traversal of a vertical link has no effect on the value of an accumulated sum.

The fast sequential summation algorithms can be developed using the observation that the line segments at high levels of this data structure provide the total length for large numbers of consecutive line segments at the deepest level. Thus, using high levels would be beneficial for the performance of such algorithms. Figure 4 illustrates the idea of the top down processing that finds a sum of the first consecutive elements. For comparison, this figure also shows the sequential summation of elements in a basic linked list.

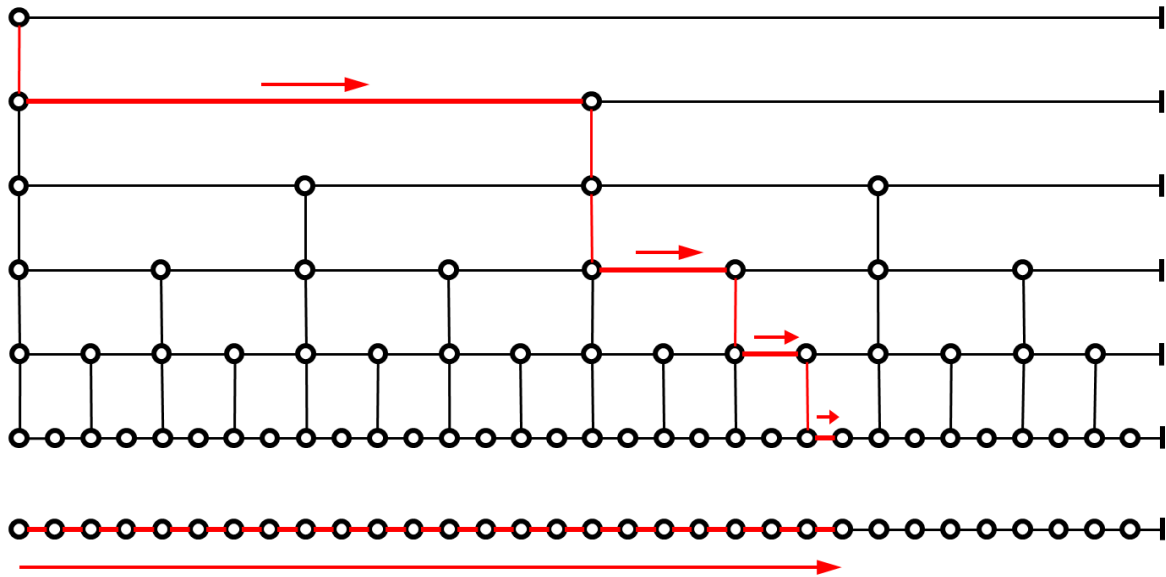

**Figure 4:** A geometric illustration of the idea of efficient sequential summation using a B+ tree with precomputed data. A sum of first elements can be computed by top down traversal of the tree. The path drawn in red color has 4 horizontal line segments with the total length equal to the total length of 23 line segments in a linked list shown at the bottom.

The optimal implementation of the top down algorithm can be expected to find the minimum possible number of line segments that cover the corresponding polyline of the deepest level. The algorithm visits all of the levels of the data structure. The number of levels is logarithmic in the number of sequence elements. The number of nodes visited at each level does not exceed the maximum size of a single group. Thus, it can be expected that the proposed data structure supports an algorithm that computes the sum of any number of first consecutive elements of a sequence in logarithmic time.

## 5. Representation of Sequence

The proper implementation of summation algorithms based on the ideas discussed so far faces the following problems. First of all, it is desirable to develop a summation algorithm that operates on any consecutive elements, not just the first ones. The second problem is that the data structure, which stores summation data only, does not yet have a method to identify the elements to be processed. Unlike basic search trees, the value or key of an element cannot be used as an identifier for the element. The summation rules are formulated for sequences, which can contain elements with multiple and unique values in sorted or random order.

A mathematical sequence is formally defined as a function that maps an element of the set of non-negative integers, which represent ranks, into an element of a given set. Each element in a sequence is uniquely identified by its rank. Note that the rank of an element is also called the subscript, the index or the position in linear order. Any consecutive elements of a sequence are defined through the ranks of the first and last elements. The formal definition requires implementing an efficient data structure that stores elements of a given set and provides access to any element specified by its rank.

The proposed data structure meets the requirements for a sequence, since it stores elements in a linked list of the deepest level. A doubly linked list also provides optimal constant time access to previous and next elements. The problem for the data structure is how to support the rank of each element with efficiency sufficient for the fast summation algorithm. The straightforward method of counting elements in a sequence using the links between the nodes in the list of the deepest level is not acceptable. This method has linear running time. It is too slow for the fast summation algorithm, which is expected to run in logarithmic time.

An interesting and probably surprising fact is that the idea of the discussed fast sequential summation algorithm using the proposed data structure hints at the solution to this problem. This conclusion follows from the observation that the rank of an element in a sequence, by definition, is the number of elements preceding this element. In other words, the rank is a special sum, which is also called the count. With the aid of the proposed data structure and the fast summation algorithm, the count and, thus, the rank can be quickly computed if each node at the deepest level stores the integer value 1 (see Figure 5).

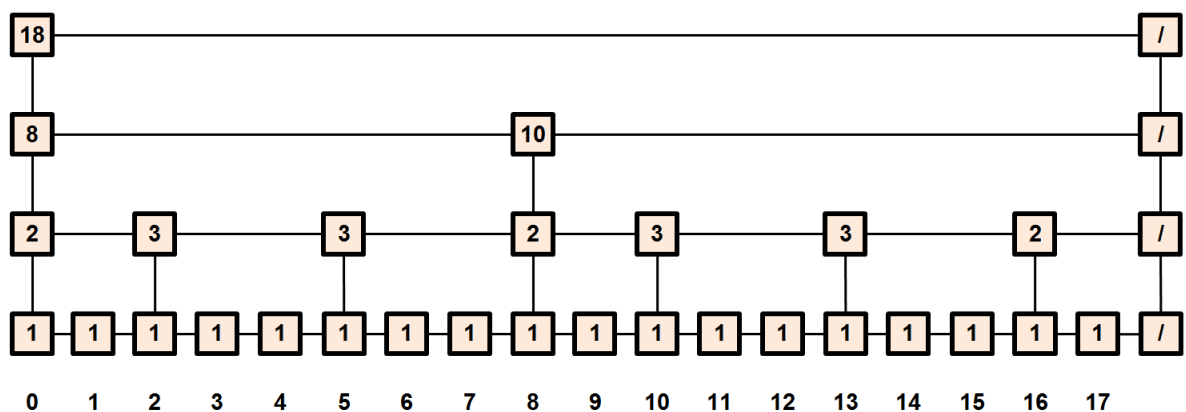

**Figure 5:** A B+ tree that stores counts of nodes at the deepest level. The bottom line shows the ranks of these nodes. Each parent node in this tree has the same number of child nodes as the matching parent node in the tree shown in Figure 2.

Due to the close similarity, the data structure of counts can be regarded as a special variant of the previously discussed data structure of sums (compare Figure 5 with Figure 2). It differs only in the type of data it stores. Its nodes store the counts of nodes at the deepest level, which are equal to the counts of sequence elements, instead of the sums of the values of elements in the same counted external nodes. At high levels, the count in a node represents the number of sequence elements in a sub-tree rooted at this node.

The algorithm for the computation of the rank of a sequence element can be implemented using the idea of the discussed fast sequential summation algorithm (see Figure 4) that operates on the data structure of counts shown in Figure 5. This algorithm represents a top down search for a node at the deepest level that stores a sequence element associated with a given rank. The algorithm uses the total count of nodes at the deepest level. As the algorithm scans through the linked list of a current level, it updates the total count using the counts stored in visited nodes. The algorithm initializes the total count to zero and starts the search at the first node at the top level. The algorithm scans forward through the current level while the total count is less than the given rank. If the current level is the deepest one, the search stops and the algorithm returns the current node. Otherwise the algorithm moves down to the next level and continues the forward scan and search.

The running time of this top down algorithm is guaranteed to be logarithmic in the number of sequence elements. It is determined by the number of levels, which is logarithmic, and the maximum size of a single group of child nodes in a B+ tree, which is constant.

The data structure that stores only the integer value 1 in each node at the deepest level is not particularly useful in practice. A general sequence stores elements of an application specific set. In order to take advantage of the efficient support for the rank and obtain a general representation of a sequence, the data structure of counts must be combined with a data structure that stores elements of the application specific set.

A B+ tree is one of the most appropriate choices for this purpose, because a tree of this type is implemented as a combination of two data structures [4]. An application data set is stored in a linear data structure that supports efficient sequential access to the data. The other data structure, which is connected to the linear structure, is a B-tree. This tree stores implementation specific data and supports fast search and random access to the elements of the application data set. The B+ trees are well-studied and offer a wide variety of options for specific implementations. The discussed data structure uses the simple combination of a level-linked B-tree with a doubly linked list.

The main focus of this discussion is on the design of fast summation algorithms. To avoid confusion, it is necessary to clarify the following aspect of the representation of a mathematical sequence. The formal definition of a sequence does not require that its elements support the addition operation. A general sequence can be used to store elements of various types, not just numerical. A classical example of a non-numeric sequence is a string of characters (see Figure 6). The summation of characters is meaningless from a human perspective.

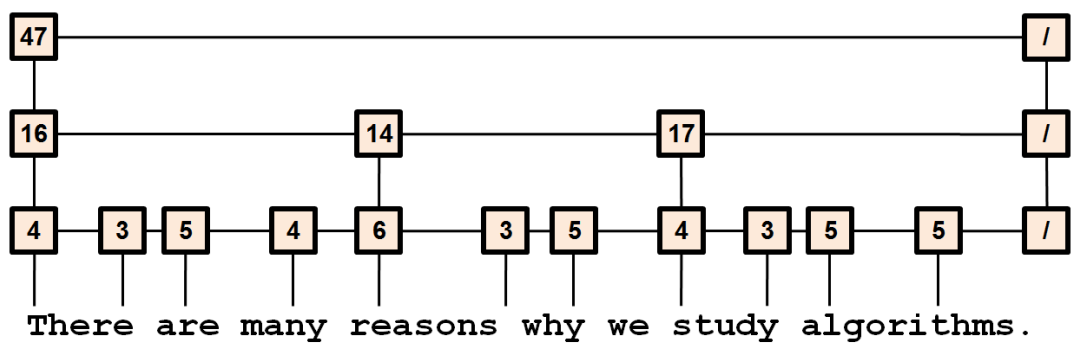

**Figure 6:** An example of a representation of a sequence of characters using a B+ tree that provides efficient sequential and random access to elements. The internal nodes store counts of characters. The counts of 1's and links between nodes at the deepest level are omitted from this figure.

The obtained data structure with efficient counting nodes at the deepest level satisfies the requirements for a mathematical sequence. The doubly linked list of the deepest level can store any data as it does a basic linked list. In addition to this, the method for the fast computation of the rank provides efficient random access to sequence elements. Normally, it should be used for distant elements, since access to adjacent elements is optimal through links to next and previous nodes. Because of these properties, the data structure that implements the discussed counting can be regarded as a linked list with improved efficiency of access to distant elements.

## 6. Fast Sequential Summation Algorithms

Efficient access to elements does not guarantee high performance of summation algorithms. If a data structure does not implement proper facilities, it can support only sequential summation, which has linear running time.

The data structure that supports a fast sequential summation algorithm is more complex than the efficient data structure, which implements a general sequence. In order to efficiently support both the summation operations and the rank method, the internal nodes of this data structure must store both sums of the values of elements in a sequence and counts of these elements. Such a data structure is a combination of the sub-structures of sums and counts (see Figure 7).

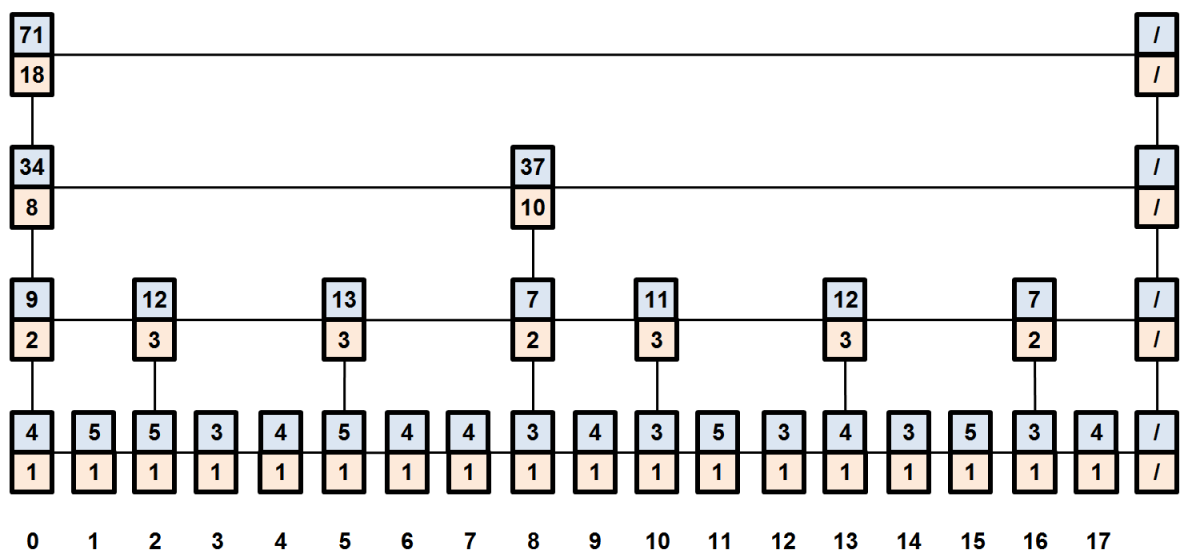

**Figure 7:** A B+ tree that provides an efficient representation of a sequence and supports fast sequential summation algorithms. The tree is a combination of the trees shown in Figures 2 and 5. Each node in this tree stores both a count and a sum of the values of elements in a sequence. All of the trees in Figures 2, 5 and 7 have an identical structure.

Despite the fact that, in theory, the sub-structure of sums supersedes the sub-structure of counts, the former does not replace the latter in a summation algorithm. These sub-structures store two different data sets and support different algorithms. The sub-structure of counts stores only non-negative integers, whereas the sub-structure of sums can store elements of any type that support the addition operation, for example, complex numbers. A fast sequential summation algorithm involves the facilities of both sub-structures. The sub-structure of counts plays an essential role, since the summation of sequence elements is specified through their ranks. Without the sub-structure of counts that supports the efficient rank method, a summation algorithm cannot efficiently identify the elements to be processed. The proposed level-linked data structure shown in Figure 7 can support two variants of efficient sequential summation algorithms that are useful in applications.

The fast sequential algorithm for a partial sum that computes the sum of the first *n* consecutive elements

$$S_n = \sum_{i=0}^{n-1} a_i \quad (1)$$

can be implemented as an extension to the efficient algorithm for the rank of an element in a sequence discussed in Section 5. This summation algorithm employs the additional variable, which represents the total sum of processed sequence elements. The algorithm starts at the first node at the top level and moves to the last node at the deepest level along a path detected by the rank algorithm. As the algorithm visits a node on the path, it uses the sum stored in the node to update the total sum.

With this partial sum algorithm, a sum of any consecutive elements in a sequence can be easily computed as the difference between the sum of the first *n* elements and the sum of the first *m* elements. This simple method offers good running time, which is logarithmic in the number of elements in a sequence, but it has a few issues. The method has excessive cost associated with the sum of the first *m* elements, which is computed twice. This computation is unnecessary to some degree, since this sum does not contribute to the result. In addition to this, there is a possibility that the sum of the first *m* elements has a huge value compared to the result value. This possibility may cause an accuracy problem in practical applications.

The optimal algorithm computes a sum of only consecutive elements specified by the lower and upper limits of the summation (see Figure 8). Its implementation avoids the unnecessary computation of a sum of the first *m* elements by processing that starts and ends at the deepest level of the data structure, which stores all of the elements in a sequence.

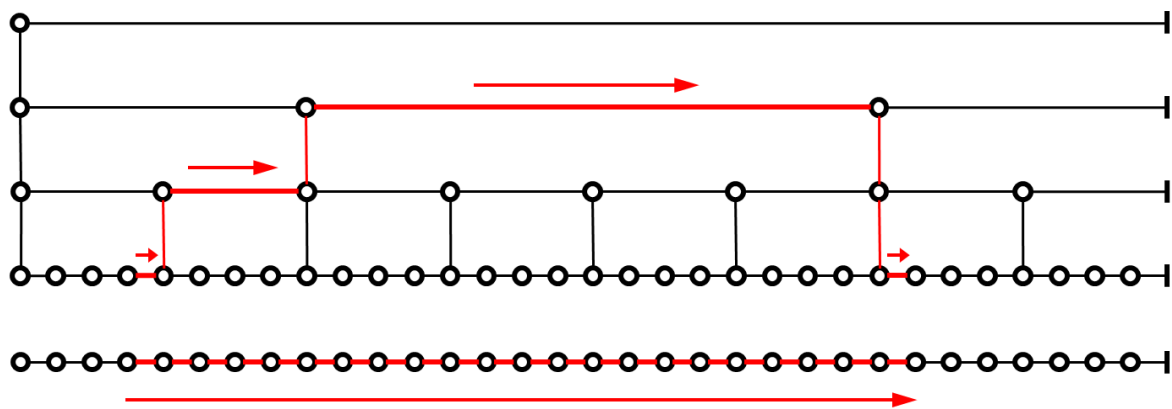

**Figure 8:** A geometric illustration of a fast sequential summation algorithm using a shortest path between two nodes at the deepest level. The algorithm visits 8 nodes (excluding the node associated with the end of the last visited line segment) at the three lowest levels and finds 4 horizontal line segments with the total length equal to the total length of 22 line segments in a linked list shown at the bottom.

This algorithm uses the given summation limits to obtain the ranks of the first and last elements to be processed. These ranks enable the algorithm to identify the nodes at the deepest level that store the first and last elements. The fast summation algorithm uses the sub-structure of counts to find a shortest path (not necessarily unique) from the first node to the last node. This implementation detail of the summation algorithm is similar to the finger search algorithm in search trees [5, 6]. The level-linked data structure provides the set of links between nodes, which is optimized for local search and allows the algorithm to use the minimum number of low levels. The algorithm reaches the top level only when the number of elements to be processed approaches the total number of elements in a sequence.

A shortest path guarantees the most efficient computation of a sum of all of the elements between two end nodes at the deepest level. As the algorithm visits nodes on the path, it uses sums stored in the nodes to compute the total sum. The summation algorithm based on a shortest path performs the minimum number of computational operations.

The summation algorithm based on finger search addresses all of the issues of the partial sum algorithm. Generally, it is more efficient, since its running time is logarithmic in the number of specified consecutive elements against the number of all of the elements in a sequence. The algorithm based on finger search does not sum up elements before the lower summation limit and, thus, provides basically the same accuracy of the result as the standard sequential summation. This advantage can be important in practice to minimize computational numerical errors. Compared to the partial sum algorithm, the summation algorithm based on finger search is more general. It can be applied to any consecutive elements in a sequence, not just the first ones.

From an implementation perspective, both summation algorithms are quite similar (see Figures 4 and 8). Each algorithm computes a sum using a shortest path from the first node to the last node. The main difference is in the choice of the first node. The partial sum algorithm starts at the first node at the top level, whereas the finger search algorithm starts at the node at the deepest level that corresponds to the lower limit of the summation. For both algorithms, a path terminates at the same last node at the deepest level that corresponds to the upper limit of the summation.

Note also that search operations on counts that support the efficient rank method and shortest path algorithms are similar to search operations on keys in a search tree. The difference is that a search for a key involves comparisons only while the search operations on counts use addition and comparison with an accumulated value.

## 7. Mean Value and Standard Deviation

The summation algorithms with logarithmic running time can significantly improve the efficiency of solutions in numerous application areas, such as numerical integration, statistics, data analysis, etc. For example, the computation of important statistical parameters, such as the mean value and the standard deviation, using normal sequential summation takes linear time. With the fast summation algorithms, these parameters can be computed in logarithmic time only. This result follows from the fact that the running times of the computation of both parameters are determined by summation operations. The number of involved elements is calculated in constant time through the specified summation limits or the ranks of corresponding elements. This operation has no significant effect on the total running time.

The mean value of any consecutive elements is simply computed by dividing an obtained sum by the number of elements

$$\frac{1}{n-m} \sum_{i=m}^{n-1} a_i \quad (2)$$

The standard deviation can be computed simultaneously with the mean value. The straightforward method is to create a new augmented data structure, which implements the fast summation of squared values of sequence elements. The more advanced method is to store the sums of squared values of elements in the nodes of the available data structure that implements a fast summation algorithm. These additional data enable the summation of squared values with the same efficiency as the summation of values of specified consecutive elements. The standard deviation is computed by taking the square root of the variance

$$\left(\frac{1}{n-m}\sum_{i=m}^{n-1} a_i^2\right) - \left(\frac{1}{n-m}\sum_{i=m}^{n-1} a_i\right)^2 \qquad (3)$$

## 8. Other Data Structures

Efficient representations of a mathematical sequence and sequential summation algorithms with logarithmic running time can be devised with various data structures. Several well-known and widely used data structures are briefly considered here.

The proposed data structures and algorithms have been developed using the method of augmenting. A general explanation of this method is provided in [2]. The underlying data structure used in specific examples is a red-black tree. Another application of this method to binary trees is available in [7]. These books show how the method of augmenting a data structure improves the performance of order-statistic queries and range queries. There are also thorough descriptions of update operations on the augmented red-black and binary trees.

In terms of the method of augmenting, a data structure that efficiently supports a sequence implements single augmenting with counts of elements. A data structure that supports a fast sequential summation algorithm implements double augmenting with both counts and sums of element values. Every augmenting uses a basic data structure to build its own sub-structure with a specific data set. Multiple augmenting can be regarded as implementing several data structures within one. For example, the discussed fast computation of the standard deviation is achieved by triple augmenting. The method of augmenting can be applied several times to one data structure to achieve the best efficiency of an algorithm.

The augmented red-black and binary trees [2, 7] efficiently support counting and the rank of an element. These facilities can be used in algorithms for the representation of a mathematical sequence. The fast sequential summation algorithms can be obtained through augmenting that creates a sub-structure of sums of element values. The second augmenting is not expected to be challenging because of the close similarity of the sub-structures of counts and sums. The augmented red-black and binary trees have, however, one important limitation, namely that they do not provide efficient finger search, which is quite useful in practice for summation algorithms.

Another data structure suitable for implementing fast sequential summation algorithms is a randomized skip list. Its simple insertion, deletion and traversal operations significantly reduce the cost of development. In the context of implementing summation algorithms, this data structure has the important advantage that it provides both top down and finger searches. A skip list can support both variants of fast sequential summation algorithms described here, since the efficient rank and summation algorithms are based on these searches in augmented data structures.

The basic data structure used here for the description of fast sequential summation algorithms represents a level-linked B+ tree with the minimal number of child-parent links. It is equivalent to a deterministic skip list [8, 9]. The methods of augmenting this basic data structure and a randomized skip list are identical. For augmenting a skip list with the counts of elements, see [10].

The benefit of the low cost of the development of an augmented skip list comes at the expense of the disadvantages that might result from the randomization of its structure and loss of balance. A skip list guarantees logarithmic running time of the discussed algorithms on average only. In addition to this, random deviations from a shortest path may lead to a loss in accuracy of the summation result.